\documentclass[superscriptaddress,amsmath,amssymb,aps,prb,twocolumn]{revtex4-2}
\usepackage[utf8]{inputenc}
\usepackage[T1]{fontenc}
\usepackage{comment}
\usepackage{amsmath}
\usepackage{amsfonts}
\usepackage{amssymb}
\usepackage{bm}
\usepackage{braket}
\usepackage{graphicx}
\usepackage[dvipsnames]{xcolor}
\usepackage[normalem]{ulem}
\usepackage{comment} 
\usepackage{hyperref}
\usepackage{algpseudocode}
\usepackage[ruled,vlined,linesnumbered]{algorithm2e}

\begin{document}

\title{Orthogonal Quantum Krylov Diagonalisation}

\author{Hadi Rammal}
\affiliation{Universit\'e de Bordeaux, CNRS, LOMA, UMR 5798, F-33400 Talence, France}

\author{Alexandre Perrin}
\affiliation{Universit\'e de Bordeaux, CNRS, LOMA, UMR 5798, F-33400 Talence, France}

\author{Oumaya Ladhari}
\affiliation{Universit\'e de Bordeaux, CNRS, LOMA, UMR 5798, F-33400 Talence, France}
\affiliation{TotalEnergies, 2 place Jean Millier, 92078 Paris La Défense Cedex, France}

\author{Clément Dutreix}
\affiliation{Universit\'e de Bordeaux, CNRS, LOMA, UMR 5798, F-33400 Talence, France}

\author{Jérémie Messud}
\affiliation{TotalEnergies, 2 place Jean Millier, 92078 Paris La Défense Cedex, France}

\author{Matthieu Saubanere}
\affiliation{Universit\'e de Bordeaux, CNRS, LOMA, UMR 5798, F-33400 Talence, France}

\begin{abstract}

Quantum subspace-diagonalization methods, particularly Quantum Krylov Diagonalization (QKD), provide a promising route for computing low-energy spectra of quantum many-body Hamiltonians. However, existing quantum Krylov approaches rely on non-orthogonal Krylov bases, requiring overlap-matrix regularization that limits numerical stability and accuracy. In this work, we introduce an Orthogonal Quantum Krylov Diagonalization (OQKD) framework that reformulates the classical Lanczos recursion at the operator level, enabling an orthogonal quantum implementation of Krylov-subspace diagonalization. By expressing Lanczos vectors as polynomial transformations of the Hamiltonian, OQKD reproduces the orthogonality, tridiagonal structure, and convergence behavior of the classical Lanczos algorithm thus eliminating the need for overlap-matrix regularization. We further show that the required Lanczos polynomials can be implemented using block encoding and Generalized Quantum Signal Processing with the same asymptotic query complexity as Chebyshev-based QKD methods. Numerical simulations of the $J_1$--$J_2$ Heisenberg model confirm the classical Lanczos convergence and numerical stability of the proposed method, while the measurement-complexity scaling is established analytically.
Building upon the OQKD framework, we then introduce a restarted state-preparation protocol that replaces a single high-degree polynomial transformation with a sequence of fixed low-degree transformations, maintaining an affordable block encoding success probability while retaining comparable convergence. These results establish OQKD as an orthogonal quantum analog of the classical Lanczos algorithm and identify the restarted protocol as a promising state-preparation strategy for Quantum Phase Estimation.

\end{abstract}

\maketitle

\section{Introduction}
The estimation of ground-state and low-lying excited-state energies of quantum Hamiltonians has emerged as a central target for quantum computing. Quantum approaches aim to alleviate the exponential computational cost associated with storing and manipulating many-body wave functions and Hamiltonian matrices on classical computers. 

Existing quantum algorithms for eigenvalue estimation can broadly be divided into three major classes: quantum phase estimation (QPE)~\cite{kitaev1995}, variational quantum algorithms (VQAs), and subspace-diagonalization approaches~\cite{yoshioka2025krylov,motta2024subspace,oumarou2026}. QPE constitutes the most direct route toward fault-tolerant quantum advantage, providing asymptotically exact eigenvalue estimation through controlled time evolution and phase extraction~\cite{Wan2021RandomizedPE,Ding2024QMEGS,OBrian2019MultipleEigenQPE,Ni2023LowDepthQPE,Kitaev2002,Aharonov1997}. Yet QPE faces a critical challenge known as the orthogonality catastrophe, where a small overlap of the initial state with the target eigenstate drastically affects the success probability restraining the conditions for a potential quantum advantage~\cite{lee2023evaluating,hpt6-9tnk,Chan2012,McClean2014}. Variational quantum algorithms, most notably the variational quantum eigensolver (VQE), were introduced as hardware-efficient approaches suitable for near-term (or NISQ) quantum processors. These methods rely on the variational optimization of parametrized quantum circuits within a classical feedback loop ~\cite{Peruzzo2014,preskill2018nisq,cerezo2021variational,tong2022heisenberg,takagi2023complexity,GENERAL_MOTA,6hpp-zl5h}. Nevertheless, their performance is often hindered by optimization difficulties such as barren plateaus, noise sensitivity, and dependence on the expressibility of the chosen ansatz~\cite{mcclean2018barren,uvarov2021barren,wang2021noise,cerezo2021cost,zimboras2025myths,lee2023evaluating}.

A third, more recent class of approaches is based on subspace-diagonalization techniques, in which a set of basis states is generated and used to construct a projected Hamiltonian within a reduced subspace. Solving the resulting projected eigenvalue problem yields approximations to the low-energy spectrum of the original Hamiltonian. A particularly successful class of such methods is based on Krylov subspaces~\cite{Krylov1931}, where the basis is generated through repeated applications of the Hamiltonian to an initial reference state. In classical numerical linear algebra, orthogonalization of the Krylov basis leads to the widely used Lanczos algorithm, which provides improved numerical stability and convergence properties~\cite{saad1980lanczos}. However, constructing and manipulating Krylov subspaces classically requires the evaluation of quantities whose storage requirements and computational cost grow exponentially with system size. 

Quantum computers would offer the possibility of generating and processing these subspaces more efficiently, thereby mitigating this bottleneck. Motivated by this potential advantage, the development of quantum subspace-diagonalization algorithms has become an active area of research~\cite{motta2024subspace}. This family of methods includes quantum Krylov diagonalization (QKD)~\cite{motta2024subspace,kirby2023lanczos,yoshioka2025krylov}, quantum filter diagonalization~\cite{cohn2021qfd}, quantum imaginary-time evolution approaches~\cite{McArdle2019,Motta2020}, and quantum selected configuration interaction methods~\cite{Baek2022}. Several versions of these algorithms, including QKD  methods, have already been demonstrated on quantum hardware~\cite{yoshioka2025krylov}. Compared to variational approaches, subspace methods often exhibit more systematic convergence and avoid challenging nonlinear optimization landscapes.

Despite these advantages, many existing QKD  methods rely on approximate constructions of non-orthogonal Krylov vectors, such as truncated imaginary-time evolution or unitary approximations of non-unitary operators.~\cite{Parrish2019,Stair2020,Cohn2021,Klymko2022,Seki2021,PhysRevA.105.022417}. 
More recently, Kirby \emph{et al.}~\cite{kirby2023lanczos} proposed a quantum Krylov construction based on block encoding and qubitization of Chebyshev polynomials~\cite{Low2019}, hereafter named CQKD. While this approach significantly improves the numerical stability of standard QKD constructions, orthogonality between Krylov vectors is still not preserved. Consequently, they do not exactly reproduce the orthogonality, stability, and convergence properties of classical Lanczos-type methods. This represents an issue as the non-orthogonality between Krylov vectors discards the use of the Krylov method on a classical computer to avoid ill-conditioning of the generalized eigenvalue problem, in this case orthogonalized variants such as the Lanczos algorithm are preferred~\cite{saad1980lanczos}.
To mitigate ill-conditioning, thresholding and regularization procedures can be introduced, resulting in additional computational and measurement costs that partially offset convergence properties and can hamper a potential advantage in a quantum computing context~\cite{kirby2023lanczos}. The primary challenge in developing a quantum analog of the Lanczos algorithm stems from its recursive structure, which requires the manipulation of previously generated Krylov vectors. Developing a quantum construction that preserves the orthogonality and convergence properties of the classical Lanczos procedure, while avoiding overlap-matrix regularization and quantum-memory requirements, therefore remains an outstanding challenge.

In this work, we introduce an Orthogonal Quantum Krylov Diagonalization (OQKD) framework that reformulates the classical Lanczos recursion at the operator level, enabling an orthogonal quantum implementation of Krylov-subspace diagonalization based on block-encoding and General Quantum Signal Processing (GQSP)~\cite{motlagh2024gqsp,sunderhauf2023gqsvt}. The resulting approach reproduces the convergence properties of the classical Lanczos algorithm while eliminating the overlap-matrix regularization required in existing quantum Krylov methods. Additionally, it requires only measuring diagonal elements of operators, avoiding the use of the costly Hadamard-tests, and thanks to the tridiagonal form of the projected Hamiltonian in the Lanczos subspace, the number of terms to be measured is reduced by a polynomial order. Moreover, the estimated ground state vector in the Lanczos subspace can also be efficiently constructed using GQSP, opening an avenue for state preparation. It also enables the introduction of a restarted variant of OQKD that replaces a single high-degree polynomial transformation with a sequence of fixed low-degree transformations, thereby maintaining a reasonable probability of successful Krylov state preparation while retaining convergence behavior comparable to the original OQKD algorithm. More broadly, this work establishes a direct connection between classical Lanczos subspace-diagonalization techniques, quantum algorithms based on block encoding and GQSP, providing an efficient quantum solver and efficient state preparation algorithm.

The remainder of the paper is organized as follows. We first review existing quantum Krylov methods and their connection to the classical Lanczos algorithm. We then present the OQKD framework together with its implementation using block encoding and GQSP. Next, we benchmark OQKD on the $J_1$--$J_2$ Heisenberg model and analyze its convergence properties together with the asymptotic scaling of its measurement complexity. Building on these results, we introduce a restarted state-preparation protocol based on OQKD and demonstrate that it enables the preparation of high-fidelity initial states with a nearly constant probability of success, making it a promising state-preparation strategy for Quantum Phase Estimation (QPE).

\begin{figure}[t]
    \centering
    \includegraphics[width=\columnwidth]{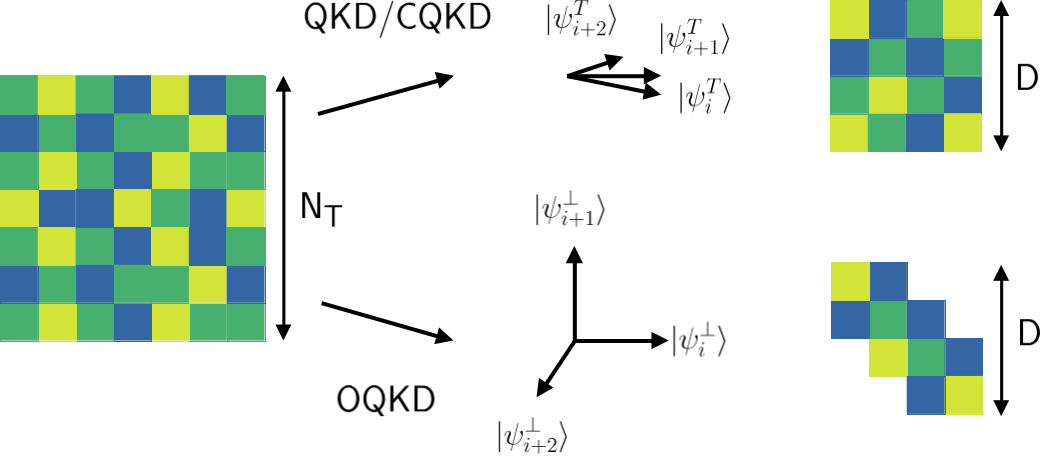}
    \caption{Schematic comparison between conventional QKD/CQKD and the proposed OQKD approach. In QKD/CQKD, the $N_T \times N_T$ Hamiltonian is projected onto a Krylov subspace of dimension $D$ spanned by generally nonorthogonal vectors, which are additionally unnormalized in block-encoding-based implementations. Consequently, a generalized eigenvalue problem involving an ill-conditioned overlap matrix must be solved. In contrast, OQKD projects the Hamiltonian onto an orthonormal Krylov basis, eliminating overlap-matrix regularization and improving numerical stability.}
    \label{fig:sketch}
\end{figure}

\section{Quantum Krylov Diagonalization and the ill-conditioning problem}

We briefly review the connection between  QKD, CQKD and the classical Lanczos method. Starting the Krylov iterative scheme with an initial state $\Phi$ and a Hamiltonian \(H\), the Krylov subspace of dimension \(D\) is defined as
\begin{equation}
\mathcal K_D(H,|\Phi_0\rangle)
=
\mathrm{span}\left\{
|\Phi_0\rangle,H|\Phi_0\rangle,\ldots,H^{D-1}|\Phi_0\rangle
\right\}.
\label{eq:krylov}
\end{equation}
Within a generally non-orthogonal basis \(\{|\psi_i^{K}\rangle\}\) where \( |\psi_i^{K}\rangle =\frac{1}{\|H^{i}|\Phi_0\rangle\|}H^{i}|\Phi_0\rangle\), one constructs the projected Hamiltonian and overlap matrices,
\begin{equation}
    H_{ij}^{K}=\langle \psi_i^{K}|H|\psi_j^{K}\rangle,
    \qquad
    S_{ij}^{K}=\langle \psi_i^{K}|\psi_j^{K}\rangle,
\end{equation}
and obtains Ritz approximations to the eigenvalues by solving the generalized eigenvalue problem
\begin{equation}
H^{K}\mathbf{v} = E S^{K}\mathbf{v} 
\label{eq:gevp}
\end{equation}
where $\mathbf{v}$ is an eigenvector of $H^{K}$ with eigenvalue $E$.
Although this procedure is exact within the chosen subspace, its numerical stability is governed by the conditioning of the overlap matrix \(S^{K}\). 
As illustrated in Fig.~(\ref{fig:sketch}), when the subspace dimension increases, the Krylov vectors tend to become progressively linearly dependent ($|\langle\psi_i^{K}|\psi_{i+1}^{K}\rangle|$ tends to increase towards $1$), causing the smallest eigenvalues of \(S\) to approach numerical precision. Consequently, beyond a critical dimension \(D_c\), which depends on the underlying problem and numerical precision, the generalized eigenvalue problem in Eq.~(\ref{eq:gevp}) becomes severely ill-conditioned. Alternative quantum subspace methods, such as Q-SENSE, exploit problem-specific symmetries to improve the structure of the variational basis, but their applicability is generally limited to systems possessing the corresponding symmetry constraints~\cite{patel2025qsense}.

A quantum Krylov and Lanczos-inspired approach reformulates the subspace construction in terms of polynomial transformations of the Hamiltonian acting on a fixed initial state. To the best of our knowledge, this is the most efficient quantum Lanczos-inspired implementation, as proposed in Ref.~\cite{kirby2023lanczos}. The Krylov subspace is generated using Chebyshev polynomial expansions,
\begin{equation}
    |\psi_i^{T}\rangle = \frac{1}{||T_i(\tilde H)|\Phi_0\rangle||}T_i(\tilde H)|\Phi_0\rangle,
\end{equation}

with \(\tilde H\) a rescaled Hamiltonian whose spectrum lies in the interval \([-1,1]\),

\begin{equation}
    \tilde H=\frac{H}{\alpha}, \quad \alpha\ge \| H\|
\end{equation}
where the spectral norm $\| \cdot\|$ is used. Chebyshev polynomials are particularly attractive in quantum algorithm development nowadays as they satisfy a stable recurrence relation and can be efficiently implemented within block-encoding and quantum signal-processing frameworks~\cite{sunderhauf2023gqsvt}. Moreover, the circuit depth required to implement a degree-\(D\) polynomial scales linearly with the polynomial degree~\cite{kirby2023lanczos}. Since the Chebyshev polynomials form a complete basis for bounded-degree polynomial expansions, the resulting vectors span the same Krylov subspace as in Eq. (\ref{eq:krylov}) in exact arithmetic. However, the generated Krylov basis vectors $|\psi_i^{T}\rangle$, are not orthonormal, $S^{T}_{ij}=\langle \psi_i^{T}|\psi_j^{T}\rangle\neq \delta_{ij}$, and still require solving a generalized eigenvalue problem involving a nontrivial overlap matrix. Although regularization of \(S^{T}\) improves numerical stability, it does not reproduce the exact Lanczos orthogonality condition \(S^{\perp}=I\). The numerical complexity of this method scales with the cost to compute the expectation values $H^{T}_{ij}$ and $S^{T}_{ij}$. 

The conditioning of the general eigenvalue problem in QKD and CQKD is exemplified in Fig.~(\ref{fig:Energy_error}) in the context of the \(J_1\)-\(J_2\) spin model (\(J_1=1\), \(J_2=0.5\)). As shown in the top panel, the condition number $\kappa(S)$ of the overlap matrix, defined as
\begin{equation}
\kappa(S)=\frac{\lambda_{\max}}{\lambda_{\min}},
\label{eq:conditioning}
\end{equation}
where $\lambda_{\max}$ ($\lambda_{\min}$) stands for the highest (lowest) eigenvalue of $S$, respectively. $\kappa$ measures the distance from a well-conditioned problem ($\kappa(S) = 1$) and shows an exponential increase with the number of iterations. Consequently thresholding is required to avoid numerical breakdown after 21 (23) iterations for QKD (CQKD) without thresholding, respectively. In addition to the consequent measurement overhead required for the thresholding process, and evaluated in Ref.~\cite{kirby2023lanczos}, it also affects the energy convergence as shown in the bottom panel of Fig.~(\ref{fig:Energy_error}). This effect, expected to be enhanced as the size of the system increases, questions the scalability of the quantum Krylov approaches at large scale. 

On a classical computer, the ill-conditioning problem discards the use of the Krylov method as the Lanczos algorithm bypasses this issue by constructing {\it on the fly} an orthonormal basis of the same Krylov subspace, adding a Gram-Schmidt rationalization step that leads to the three-term recurrence
\begin{equation}
H|\psi_i^{\perp}\rangle
=
\beta_{i+1}|\psi_{i+1}^{\perp}\rangle
+
\alpha_i|\psi_i^{\perp}\rangle
+
\beta_i|\psi_{i-1}^{\perp}\rangle ,
\label{eq:lanczos}
\end{equation}
with
\[
\alpha_i
=
\langle \psi_i^{\perp}|H|\psi_i^{\perp}\rangle,
\quad
\beta_{i+1}
=
\left\|
H|\psi_i^{\perp}\rangle
-
\alpha_i|\psi_i^{\perp}\rangle
-
\beta_i|\psi_{i-1}^{\perp}\rangle
\right\|.
\]
and where the overlap matrix satisfies
\begin{equation}
    S^{\perp}_{ij}=\langle \psi_i^{\perp}|\psi_j^{\perp}\rangle=\delta_{ij},
\end{equation}
such that the basis vectors are orthonormal.
In exact arithmetic, the recurrence relation in Eq.~(\ref{eq:lanczos}) produces an orthonormal basis, so that \(S^{\perp}=I\), and the Hamiltonian is represented by a tridiagonal matrix. Still in exact arithmetic, the standard Krylov and Lanczos formulations generate different basis vectors, but these vectors span the same Krylov subspace. Consequently, they yield identical Ritz values at a fixed subspace dimension~\cite{saad2011numerical}.
Their practical difference is numerical: the Lanczos procedure can stably access significantly larger effective Krylov dimensions because orthogonality is enforced by construction. Consequently, the maximal stable subspace dimension typically satisfies $D^{\mathrm{Lanczos}}_c > D^{\mathrm{Krylov}}_c$, allowing the Lanczos method to achieve substantially more accurate approximations of the extremal eigenvalues.

Despite the significant advances represented by CQKD, a gap still exists between the most advanced quantum Krylov methods and the classical Lanczos algorithm. Bridging this gap by constructing a quantum method that preserves orthogonality by construction, avoids overlap-matrix regularization, and does not require quantum memory constitutes the central challenge addressed in this work.

\section{Orthogonal Quantum Krylov Diagonalization}

\subsection{General framework}

In this section, we propose the Orthogonal Quantum Krylov Diagonalization method (OQKD) using a reformulation of the Lanczos recursion at the operator level. Following the Lanczos algorithm, each vector is written as a polynomial of the rescaled Hamiltonian acting on the initial state,
\begin{equation}
|\psi^{\perp}_n\rangle = P_n(\tilde H)|\Phi_0\rangle,
\label{eq:OQKD}
\end{equation}
normalized by construction, i.e. we always have $\|P_n(\tilde H)|\Phi_0\rangle\|^2=1$.
The polynomials \(P_n(\cdot)\) obey the rescaled Lanczos recurrence
\begin{equation}
\tilde\beta_{n+1}P_{n+1}(x)
=
(x-\tilde\alpha_n)P_n(x)
-
\tilde\beta_n P_{n-1}(x),
\label{eq:poly_lanczos}
\end{equation}
with
\[
P_0(x)=1,
\qquad
P_1(x)=\frac{x-\tilde\alpha_0}{\tilde\beta_1}.
\]
Here, we adopt a generic normalization of the Hamiltonian spectrum. The following derivation remains valid for any normalization scheme of the form
\begin{equation}
    \tilde H=\frac{H}{\alpha}, \quad \alpha\ge \| H\|
\end{equation}
with the corresponding Lanczos coefficients transformed as
\begin{equation}
    \tilde\alpha_n=\frac{\alpha_n}{\alpha},
\qquad
\tilde\beta_n=\frac{\beta_n}{\alpha}.
\end{equation}
Once the Lanczos coefficients are known, the full Lanczos basis can be generated by applying the corresponding polynomial transformations to the same initial state.

To complete the recursive construction, one must determine the normalization coefficients \(\tilde\beta_{i+1}\), which are required to generate \(|\psi^{\perp}_{i+1}\rangle\) without explicitly storing intermediate Lanczos vectors as it is done classically. This can be achieved recursively from measurable moments. Indeed, using the tridiagonal structure of \(H\) in the Lanczos basis, one obtains
\begin{equation}
    \langle \psi^{\perp}_i|\tilde H^2|\psi^{\perp}_i\rangle = \tilde\alpha_i^2+\tilde\beta_{i+1}^2+\tilde\beta_i^2,
\end{equation}
and therefore
\begin{equation}
\tilde\beta_{i+1}^2
=
\langle \psi^{\perp}_i|\tilde H^2|\psi^{\perp}_i\rangle
-
\tilde\alpha_i^2
-
\tilde\beta_i^2,
\label{eq:beta_recursion}
\end{equation}
leading to the aforementioned inherent normalization in Eq. (\ref{eq:OQKD}).
The derivation of Eq.~(\ref{eq:beta_recursion}) is provided in the Appendix~\ref{app:derivation}. The quantities on the right-hand side are either known from previous iterations or can be estimated directly on the quantum processor. This relation closes the recursion and allows the Lanczos polynomials to be constructed iteratively. Note that in the standard classical Lanczos method $\{\tilde{\beta}_i\}$ are obtained {\it a posteriori} from the construction of the Lanczos vector as a non-diagonal element of the Hamiltonian in the reduced basis, or equivalently, to preserve the norm of the Lanczos vector. Here, $\{\tilde{\beta}_i\}$ are obtained {\it a priori} to determine the coefficients of the Lanczos polynomial. Moreover, only the sets $\{\tilde{\alpha}_i\}$ and $\{\tilde{\beta}_i\}$ need to be measured at each iteration, and thanks to Eq.~\ref{eq:beta_recursion} do not require any off-diagonal element of the Hamiltonian nor overlap measurement, thus bypassing the use of the Hadamard test.  For completeness, we note that an exact \emph{happy breakdown}, characterized by $\tilde{\beta}_n=0$, is a rare event in finite-precision arithmetic. In this case, the Krylov subspace is invariant under the Hamiltonian, the Ritz eigenpairs are exact, and the iteration terminates.

\begin{figure}[t]
    \centering
    \includegraphics[width=\columnwidth]{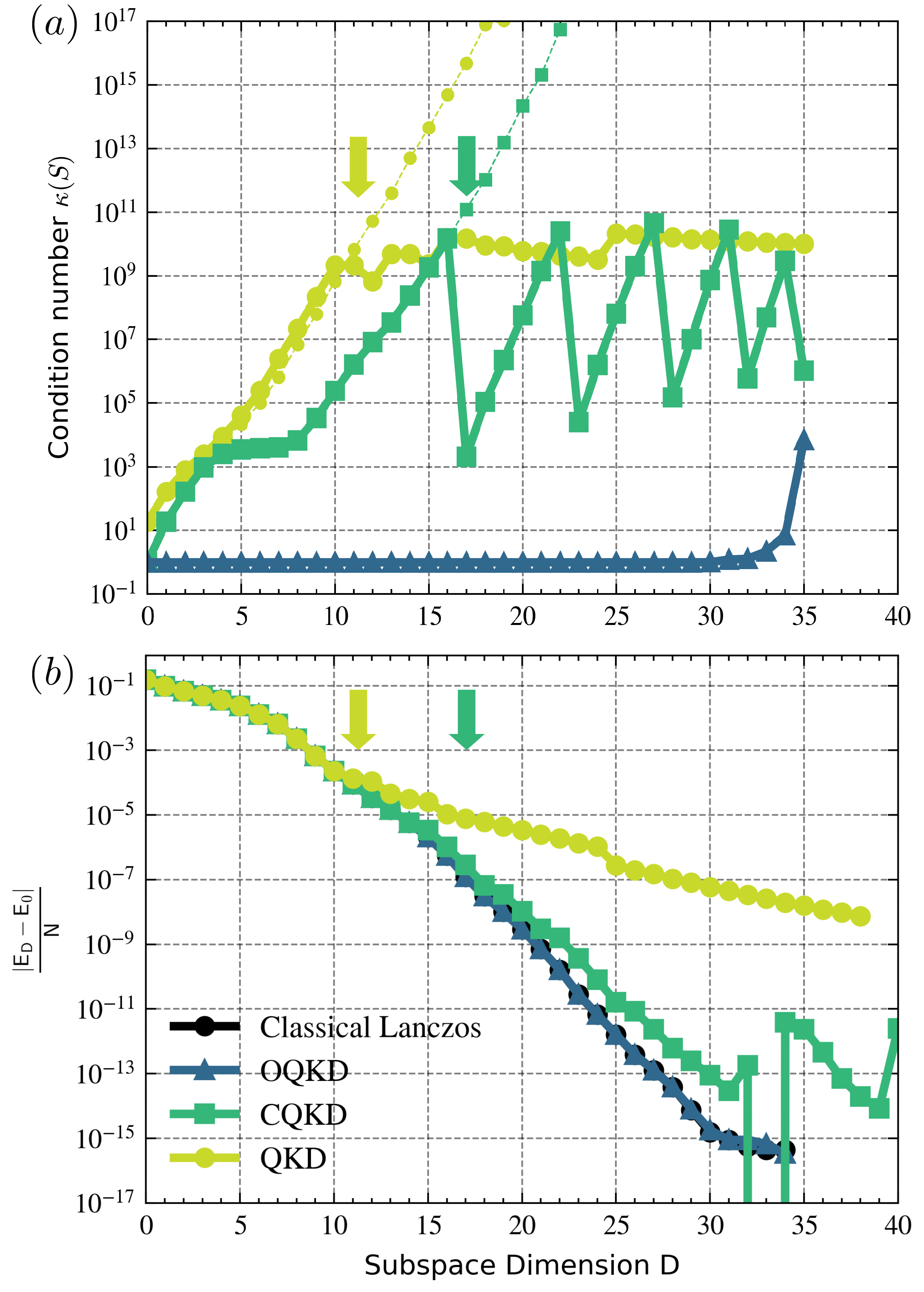}
    \caption{
Results for the \(J_1\)-\(J_2\) spin model (\(J_1=1\), \(J_2=0.5\)) with an initial mean-field approximated ground state, comparing QKD, CQKD, and OQKD. The arrows indicate the first subspace dimension at which the thresholded and unthresholded calculations begin to differ.
(a) Condition number of the overlap matrix \(S\) as a function of Krylov-subspace dimension. QKD becomes increasingly ill-conditioned with increasing subspace dimension, while CQKD partially mitigates this effect through the Chebyshev thresholding technique. In contrast, OQKD preserves orthogonality by construction, resulting in an overlap matrix that remains close to the identity up to numerical precision. The dashed curves indicate the corresponding results obtained without thresholding.
(b) Ground-state energy error as a function of Krylov-subspace dimension. OQKD reproduces the convergence behavior of the classical Lanczos algorithm up to numerical precision and reaches machine precision, whereas overlap-conditioning effects lead to slower convergence in QKD and CQKD.
The arrows indicate the first Krylov subspace dimension at which thresholding is activated. In both QKD and CQKD, a threshold of $10^{-9}$ was employed.}
    \label{fig:Energy_error}
\end{figure}

\subsection{Practical implementation}
\label{parag:impelntation}

Following the definition of Lanczos polynomials $P_n(\cdot)$ in Eq. (\ref{eq:OQKD}),
we describe a quantum implementation of their application to $\tilde H$ and $|\Phi_0\rangle$ based on block encoding. The Lanczos polynomial is expanded in the Chebyshev basis, which is particularly well suited for our implementation. Indeed, Chebyshev polynomials constitute the natural polynomial basis for quantum signal
processing and block-encoding techniques, and their simple recurrence and multiplication relations enable an efficient manipulation of polynomial transformations. Accordingly, we write
\begin{equation}
P_n(x)
=
\sum_{k=0}^{n}
c_{nk}T_k(x),
\label{eq:lanczos_cheb_expansion}
\end{equation}
where the coefficients \(c_{nk}\) are determined recursively from the Lanczos coefficients \(\{\tilde\alpha_k\}\) and \(\{\tilde\beta_k\}\). Details of the coefficient construction are provided in Appendix~\ref{app:C_coefficient_algo}.

We start from a block encoding of $\tilde H$
\begin{equation}
(\langle 0|_a\otimes I_s)\,
U\,
(|0\rangle_a\otimes I_s)
=
\tilde H,
\label{eq:block_encoding}
\end{equation}
where $a$ represents the "ancilla" qubits space ($A$ qubits) and $s$ the "physical system" qubits space ($S$ qubits required to describe the states $\ket{\psi_n^{\perp}}$).
There are various ways to define $U$ (together with $A$). 
In practice, as electronic systems Hamiltonians naturally take a Linear Combination of Unitaries (LCU) form with $T$ terms, $U$ is usually computed from the "PREP" $P$ and "SELECT" $S$ operators, $U=(P^\dagger\otimes I_s)S(P \otimes I_s)$,
which requires $A=\lceil\log_2 T\rceil$ ancilla qubits~\cite{Childs2017,Low2019}. We then define the qubitization walk operator
\begin{equation}
W=RU,
\end{equation}
where \(R\) is a reflection about the $|0\rangle_a$ state
in the ancilla qubits space, $R=(2|0\rangle_a\langle 0|_a -I_a)\otimes I_s$.

To implement a polynomial transformation of the Hermitian operator $\tilde H$ on a quantum computer, the GQSP framework can be used, giving:
\begin{equation}
(\langle 0|_a\otimes I_s)\,
\text{GQSP}_n[W]\,
(|0\rangle_a\otimes I_s)
=
\frac{1}{\Lambda_n}P_n(\tilde H),
\label{eq:block_encoding2}
\end{equation}
where $\Lambda_n$ is a normalization factor 
defined by
\begin{equation}
\Lambda_n\ge \max_{|z|=1}|p_n(z)|,
\quad p_n(z)=\sum_{k=0}^n c_{nk} z^k.
\label{eq:block_encoding3}
\end{equation}
Thus, $\Lambda_n=1$ can be chosen if $\max_{|z|=1}|p_n(z)|\le 1$ and $\Lambda_n>1$ must be taken if $\max_{|z|=1}|p_n(z)|> 1$.

The GQSP sequence consists of alternating applications of the walk operator and rotations,
\begin{equation}
\mathrm{GQSP}_n[W]
=
R_nWR_{n-1}W\cdots WR_1WR_0,
\label{eq:gqsp_sequence}
\end{equation}
where each rotation \(R_j\) depends on the parameters
\(\{\theta_j,\phi_j,\lambda\}\) and acts on the signal-processing ancilla qubit. The GQSP construction requires only one additional ancilla qubit beyond those used for the block encoding of \(H\), while increasing the polynomial degree affects only the circuit depth.
The rotation parameters are determined classically from the polynomial $\frac{1}{\Lambda_n}p_n(\cdot)$, i.e. from its coefficients
\begin{equation}
\frac{c_{nk}}{\Lambda_n},
\end{equation}
through the complementary-polynomial construction and recursive phase-factor decomposition as in Ref.~\cite{motlagh2024gqsp}, after which the rotation angles are extracted recursively in \(\mathcal{O}(n)\), where \(n\) is the degree of the synthesized polynomial. A brief summary of this procedure is provided in Appendix~\ref{app:angels}. For completeness, we mention that the GQSP circuit acting on the walk operator
can be viewed as a generalized quantum eigenvalue-transform (GQET) circuit directly associated with the block encoding \(U\) 
~\cite{sunderhauf2023gqsvt},
$
\mathrm{GQSP}_n[W]
=
\mathrm{GQET}_n[U].
$
Details are presented in Appendix \ref{App:GQSP}.

For the Lanczos polynomial in Eq.~(\ref{eq:lanczos_cheb_expansion}),
the GQSP boundedness condition $\max_{|z|=1}|p_n(z)|\le 1$, Eq.~(\ref{eq:block_encoding3}), is generally not satisfied. This means that a normalization factor $\Lambda_n>1$ has to be considered,
for which a lower bound is provided by eq. (\ref{eq:block_encoding3}).
This equation can be solved numerically to define the smallest possible $\Lambda_n$. Another staightworward option is to choose a larger $\Lambda_n$ through (using triangular inequality):
\begin{equation}
\Lambda_n
=
\sum_{k=0}^{n}
|c_{nk}|
\quad\ge\quad\max_{|z|=1}|p_n(z)|
.
\end{equation}
Consequently, the GQSP circuit block encodes
\begin{equation}
\frac{1}{\Lambda_n}P_n(\tilde H),
\end{equation}
and outputs, conditioned on measuring the $\lceil\log_2 T\rceil$ ancilla qubits in the $0$ state,
\begin{equation}
P_n(\tilde H)\ket{\Phi_0}
=
\ket{\psi_n^{\perp}},
\label{eq:res}
\end{equation}
as we always have $\| P_n(\tilde H)|\Phi_0\rangle\|^{2}=1$ by Lanczos construction.
The corresponding block-encoding success probability is
\begin{equation}
p_{\mathrm{succ}}^{\mathrm{OQKD}}
=
\left\|
\frac{1}{\Lambda_n}
P_n(\tilde H)
|\Phi_0\rangle
\right\|^2
=
\frac{1}{\Lambda_n^2}.
\end{equation}

Finally, the OQKD algorithm is summarized in Algorithm~\ref{app:quantum_algo}.
\begin{algorithm}[ht]
  \caption{OQKD Algorithm}\label{app:quantum_algo}
  \begin{enumerate}
  \item \textbf{Input}:
     \[
     \tilde H,\quad |\Phi_0\rangle .
     \]
  \item \textbf{Initialization}:
     \begin{enumerate}
     \item Set \({P}_0(\tilde H)=I\).
     \item Measure on a quantum
processor $\tilde\alpha_0$ and compute $\tilde\alpha_0^2$
     \item Compute $\tilde\beta_1$, see Eq. (\ref{eq:beta_recursion}).
     \end{enumerate}
  \item \textbf{Iteration}:
    For \(i=1,\ldots,N\):
     \begin{enumerate}
     \item \textbf{State preparation}:
Construct a block encoding of the Lanczos polynomial
$\frac{1}{\Lambda_i}{P}_i(\tilde H)$ using GQSP and prepare on a quantum processor
\begin{equation}
|\psi_i^{\perp}\rangle
=
{P}_i(\tilde H)|\Phi_0\rangle,
\end{equation}
see Eq.(\ref{eq:res}).
     \item \textbf{Measurement}:
       Measure on a quantum processor
       \[
       \tilde\alpha_i
       =
       \langle \psi_i^{\perp}|\tilde H|\psi_i^{\perp}\rangle ,
       \]
       and
       \[
       \langle \psi_i^{\perp}|\tilde H^2|\psi_i^{\perp}\rangle .
       \]
     \item \textbf{Normalization coefficient}:
       Compute
       \[
       \tilde\beta_{i+1}^2
       =
       \langle \psi_i^{\perp}|\tilde H^2|\psi_i^{\perp}\rangle
       -
       \tilde\alpha_i^2
       -
       \tilde\beta_i^2 .
       \]
     \item \textbf{Polynomial update}:
       Compute the coefficients of $P_{i+1}(\cdot)$ using the recursive coefficient relation (see appendix ~\ref{app:C_coefficient_algo}).
     \end{enumerate}
  \item \textbf{Classical post-processing:}
Construct the projected Hamiltonian in the Lanczos basis, recover the normalization factors associated with the Lanczos vectors, and solve the resulting projected eigenvalue problem.
  \end{enumerate}
\end{algorithm}
\subsection{Numerical assessment on $J_1$-$J_2$ model}
\label{ssec:numeric}

In this section, we present classical numerical results based on state-vector simulations performed using Eq.~\ref{eq:lanczos_cheb_expansion} using a $4\times4$ $J_1$-$J_2$ model given by
\begin{equation}
H
=
J_1
\sum_{\langle i,j\rangle}
\mathbf{S}_i\cdot\mathbf{S}_j
+
J_2
\sum_{\langle\!\langle i,j\rangle\!\rangle}
\mathbf{S}_i\cdot\mathbf{S}_j,
\label{eq:J1J2}
\end{equation}
Fig.~\ref{fig:Energy_error} compares the convergence behavior and numerical stability of different quantum subspace-diagonalization methods. 
As expected, the standard QKD method exhibits a rapid exponential growth of the condition number, indicating severe ill-conditioning of the generated Krylov basis. The CQKD construction significantly mitigates this growth and delays the onset of ill-conditioning to larger subspace dimensions. The arrows indicate the first subspace dimension at which the thresholded and unthresholded calculations begin to differ, signaling the onset of numerical instability associated with small eigenvalues of the overlap matrix. For QKD, this occurs at approximately $D \simeq 11$, while for CQKD the corresponding threshold is shifted to $D \simeq 17$. The dotted curves show the results obtained without thresholding and illustrate the point at which the unregularized overlap matrix becomes sufficiently ill-conditioned that its spectrum can no longer be reliably resolved.

Nevertheless, both methods eventually develop very large overlap-matrix condition numbers, which necessitate the regularization procedure used to stabilize the generalized eigenvalue problem. The consequences of this regularization are visible in panel~(b), which shows the corresponding ground-state energy error. Although both QKD and CQKD initially display the characteristic exponential convergence of Krylov-subspace methods, they eventually deviate from the classical Lanczos convergence profile once the overlap-matrix regularization becomes active. This deviation originates from the removal of nearly linearly dependent Krylov vectors, which effectively reduces the dimension of the projected subspace and alters the exact Lanczos recursion. For CQKD, at large subspace dimensions, the energy-error curve begins to fluctuate around a converged value determined by the chosen thresholding parameters~\cite{kirby2023lanczos}.

By contrast, the proposed OQKD method reproduces the classical Lanczos convergence behavior almost exactly, reaching machine precision with essentially the same convergence profile as the classical algorithm. In this case, the overlap matrix remains close to the identity throughout the recursion, with condition number \(\kappa \approx 1\) over the full range of arithmetically stable iterations. It is worth noting that the small deviations from unity arise from numerical arithmetic errors, which are also observed in the classical Lanczos method. In panel~(b), the OQKD curve breaks down at approximately the same Krylov-subspace dimension as the classical Lanczos algorithm, again due to the onset of numerical arithmetic limitations. 
Consequently, no overlap-matrix regularization is required, allowing the method to preserve the exact orthogonality structure of the Lanczos basis and faithfully reproduce the behavior of the classical Lanczos algorithm.

\begin{figure}
    \centering
    \includegraphics[width=\linewidth]{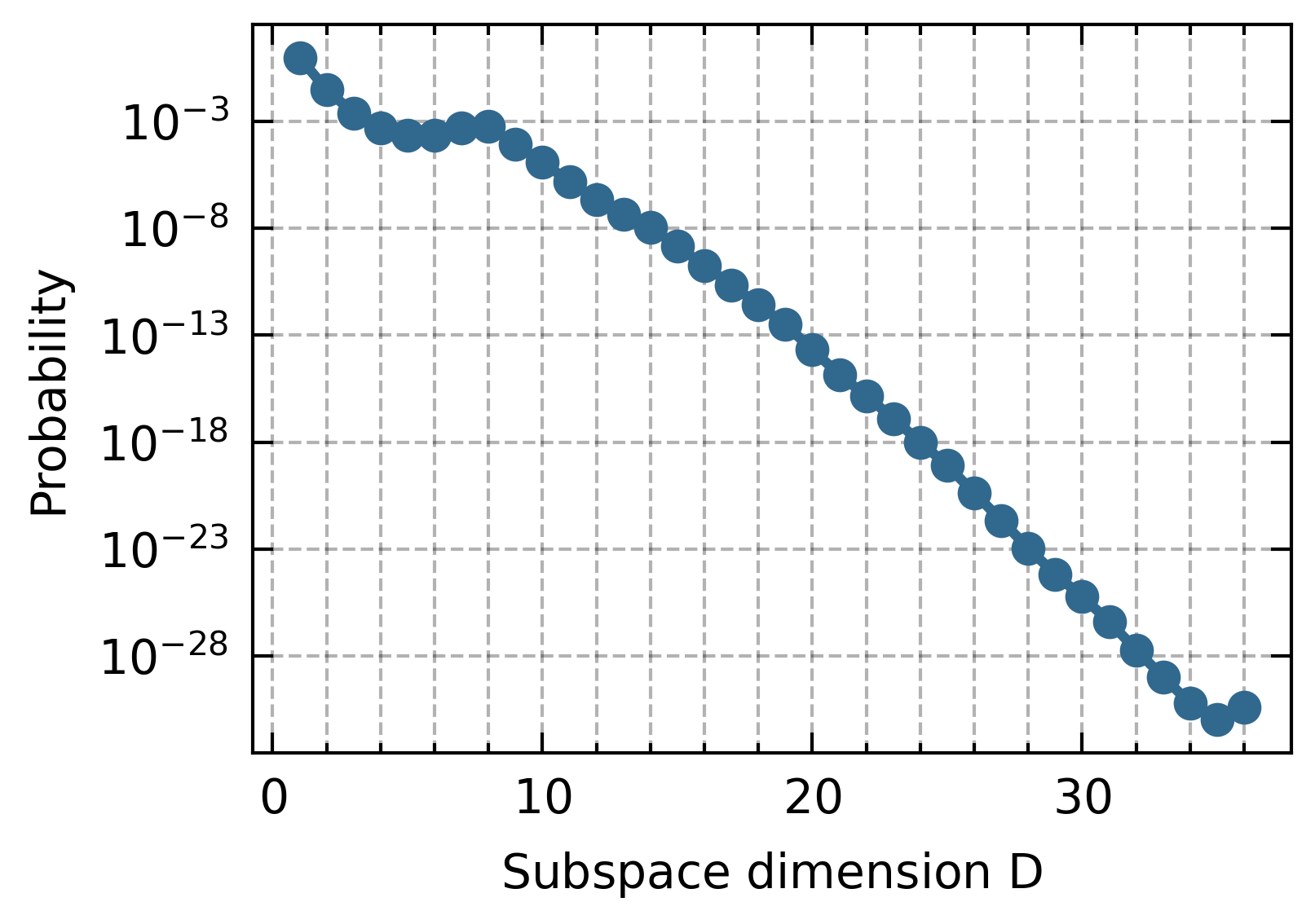}
    \caption{Probability of successfully implementing the normalized Lanczos polynomial using GQSP as a function of the Lanczos subspace dimension \(D\). The success probability decreases exponentially with increasing polynomial degree due to the rapid growth of the normalization factor \(\Lambda_n\). }
    \label{fig:P_succ}
\end{figure}

The main drawback of OKQD concerns the rapid growth of the normalization factor $\Lambda_n$ with the Lanczos dimension, causing the success probability of implementing the corresponding GQSP polynomial to decrease approximately exponentially with the polynomial degree, as shown in Fig.~\ref{fig:P_succ}. Consequently, the direct implementation of high-degree Lanczos polynomials becomes prohibitive. Note that this problem is common with GQSP approaches and can be mitigated by using the amplitude amplification method~\cite{Brassard_2002}. Importantly, this exponential overhead arises from the current implementation of polynomial transformations within the GQSP framework rather than from the OQKD algorithm itself. It therefore represents a practical implementation limitation that may be alleviated by future advances in GQSP and quantum signal-processing techniques. 

\subsection{Measurement Complexity}

One of the primary factors determining the efficiency of a quantum algorithm is its measurement cost, namely, how the number of measurements scales with the system size of the underlying Hamiltonian. In this section, we compare the asymptotic measurement complexity of OQKD with that of CQKD.

In CQKD, overlap-matrix regularization introduces an additional error contribution in the presence of statistical noise. As shown in Ref.~\cite{kirby2023lanczos}, the resulting ground-state energy error satisfies in the presence of noise sampling

\begin{equation}
\mathcal{E}
\leq
\mathcal{O}\!\left(
\frac{1}{\sqrt{M}}
+
\frac{\sqrt{\delta}}{|\gamma_0|^2\sqrt{M}}
+
\frac{1}{|\gamma_0|^2}
\left(1+\frac{\delta}{2}\right)^{-D}
\right),
\label{eq:error_bound_sampling1}
\end{equation}
where $M$ denotes the number of measurements used to estimate each projected matrix element, $\delta$ is the spectral gap, $D$ is the Lanczos subspace dimension, and
$\gamma_0=\langle E_0|\Phi_0\rangle$ is the overlap between the initial state and the exact ground state. The first term corresponds to the standard statistical sampling error, while the second originates from the thresholding procedure used to stabilize the overlap matrix. Consequently, the number of measurements required for each projected matrix element scales as

\begin{equation}
    M_{\mathrm{CQKD}} = \Theta\!\left(
\frac{1}{\mathcal{E}^{2}} + \frac{1}{\mathcal{E}|\gamma_0|^{4}}
\right).
\end{equation}
Since each measurement additionally requires successfully preparing the corresponding quantum state, the effective measurement cost becomes
\begin{equation}
    M_{\mathrm{CQKD}}^{\mathrm{eff}} = \frac{M_{\mathrm{CQKD}}} {p_{\mathrm{succ}}^{\mathrm{CQKD}}},
\end{equation}
where $p_{\mathrm{succ}}^{\mathrm{CQKD}}=||T_i(\tilde H)|\Phi_0\rangle||^2$ denotes the probability of successfully preparing the required Chebyshev state. In practice, this probability remains on the order of $10^{-1}$ for all $i$.

For many strongly correlated quantum systems, simple efficiently preparable reference states, such as Hartree--Fock product states, are expected to exhibit an exponentially decreasing overlap with the exact ground state as the system size increases, a manifestation of the orthogonality catastrophe~\cite{tubman2018postponing}. Accordingly, we assume $|\gamma_0|
=
\mathcal{O}\!\left(e^{-\alpha N}\right),
$ where \(N\) denotes the number of qubits and \(\alpha>0\) is a system-dependent constant. Consequently, the effective measurement complexity of CQKD inherits an exponential dependence on the system size.

By construction, OQKD completely avoids overlap-matrix thresholding and regularization. Under the present analysis, the overlap-dependent contribution in Eq.~(\ref{eq:error_bound_sampling1}) is therefore eliminated, leaving only the standard statistical sampling error. The measurement complexity per projected matrix element becomes

\begin{equation}
    M_{\mathrm{OQKD}} = \Theta\!\left(\frac{1}{\mathcal{E}^{2}}\right),
\end{equation}
which is independent of the initial-state overlap. However, OQKD requires the preparation of orthogonal Lanczos vectors using GQSP. Therefore, the effective measurement cost is

\begin{equation}
    M_{\mathrm{OQKD}}^{\mathrm{eff}} = \frac{M_{\mathrm{OQKD}}}{p_{\mathrm{succ}}^{\mathrm{OQKD}}(D)},
\end{equation}
where $p_{\mathrm{succ}}^{\mathrm{OQKD}}(D)$ is the probability of successfully implementing the degree-$D$ GQSP polynomial. As discussed in Sec.~\ref{ssec:numeric}, this probability decreases exponentially with the Lanczos dimension due to the growth of the polynomial coefficients.

The effective measurement complexities of the two approaches can therefore be summarized as

\begin{equation}
    M_{\mathrm{CQKD}}^{\mathrm{eff}}
=
\Theta\!\left(
\frac{1}{p_{\mathrm{succ}}^{\mathrm{CQKD}}}
\left[
\frac{1}{\mathcal{E}^{2}}
+
\frac{1}{\mathcal{E}|\gamma_0|^{4}}
\right]
\right),
\end{equation}
and
\begin{equation}
    M_{\mathrm{OQKD}}^{\mathrm{eff}}
=
\Theta\!\left(
\frac{1}{p_{\mathrm{succ}}^{\mathrm{OQKD}}(D)}
\frac{1}{\mathcal{E}^{2}}
\right).
\end{equation}

These expressions highlight the fundamental distinction between CQKD and OQKD. In CQKD, the effective measurement complexity inherits an exponential dependence on the system size through the exponentially decreasing initial-state overlap. In contrast, OQKD completely removes the overlap dependence from the measurement complexity. The remaining overhead is determined solely by the success probability of implementing the degree-$D$ GQSP polynomial, shifting the exponential dependence from an intrinsic property of the physical system to an implementation-dependent algorithmic parameter. This remaining overhead is therefore technical rather than fundamental.

\section{Restart and State-Preparation}
\label{par:ROQKD}

The direct implementation of Lanczos polynomials into quantum circuits using block encoding and GQSP paves the way for high-quality initial-state preparation. In particular, it is seems suited for Quantum Phase Estimation (QPE), whose success depends critically on the overlap between the prepared initial state and the target eigenstate.

However, the direct implementation of OQKD requires Lanczos polynomials of progressively higher degree. Their realization through GQSP becomes increasingly unlikely to succeed as the Lanczos dimension grows. This limitation originates from the systematic increase of the coefficients $c_{nm}$ appearing in the Chebyshev expansion of the Lanczos polynomials $P_n$ at each iteration $n$. 
Importantly, the diagonalization of the reduced Hamiltonian in the $n \times n$ Lanczos subspace provides, at every iteration, an approximation to the ground state, denoted by $|\Psi_0^{(n)}\rangle$ whose accuracy increases with $n$. 

At iteration $n$, the solution of the projected (generalized) eigenvalue problem provides the coefficients of the approximate ground state $|\Psi_0^{(n)}\rangle$ expressed in the Krylov basis. Since each Krylov basis vector can itself be written as a polynomial of the rescaled Hamiltonian acting on the initial state, the approximate ground state may be expressed as
\begin{equation}
|\Psi_0^{(n)}\rangle
=
Q^{(n)}(\tilde H)|\Phi_0\rangle,
\end{equation}
where $Q^{(n)}$ is the polynomial obtained by combining the Krylov basis polynomials with the ground-state coefficients. Expanding $Q^{(n)}$ in the Chebyshev basis therefore provides a direct quantum-computing-compatible representation of the approximate ground state, allowing it to be implemented using GQSP, as described below.

In contrast to the Lanczos polynomials, the Chebyshev coefficients of $Q^{(n)}$ do not exhibit significant growth with the iteration number. Since each iteration only refines the previous ground-state approximation, the associated polynomial remains well conditioned. This behavior becomes even more pronounced at large iteration numbers owing to the rapid convergence of $|\Psi_0^{(n)}\rangle$ toward the exact ground state. Consequently, the probability of successfully implementing $Q^{(n)}(\tilde H)$ through GQSP is expected to remain nearly constant throughout the iterations, despite the exponential decrease in the success probability associated with implementing the Lanczos polynomials themselves.

This observation naturally motivates a restarted OQKD protocol (ROQKD), in which the approximate ground state $|\Psi_0^{(n)}\rangle$ obtained after a given iteration is used as the new initial state for the next OQKD cycle. In this way, the algorithm avoids the exponentially decreasing success probability associated with implementing increasingly high-degree Lanczos polynomials while preserving the rapid convergence of the Krylov method. A detailed analysis of the evolution of the GQSP success probability for the accumulated restart polynomial is presented in Appendix~\ref{app:state_preparation_probability}. For completeness, we also consider a restarted Quantum Krylov protocol, in which the Krylov subspace dimension at each restart is chosen as the largest value that still permits a numerically stable solution of the generalized eigenvalue problem described in the next section. This strategy provides a direct comparison between restarted OQKD and restarted Quantum Krylov approaches.

Finally, the resulting approximate ground state $|\Psi_0^{(n)}\rangle$ can be used as the input state for complementary algorithms such as QPE, ROQKD having the potential to retain a large overlap with the target eigenstate for final enhancement.

 \subsection{Restart protocol}

In order to avoid the fast decrease of the probability of success $1/\Lambda_n^2$
to implement $P_n$, we propose to constrain the OQKD maximum number of iterations with a value $n^{\max}$. Note that equivalently we could constrain a minimum success probability through a value $\Lambda^{\rm max}$. 

The restarted algorithm is initialized as the standard CQKD or OKQD algorithm, by a trial state $|\Phi^{(k)} \rangle$, where $k$ stands for the number of restarts and is set to $0$. Following algorithm~\ref{app:quantum_algo}, the OKQD algorithm is used to prepare the $n^{\rm max}$ Lanczos polynomials

\begin{equation}
    \left\{
|\psi^{\perp(k)}_0\rangle,
|\psi^{\perp(k)}_1\rangle,
|\psi^{\perp(k)}_2\rangle,
\dots,
|\psi^{\perp(k)}_{n^{\rm max}}\rangle
\right\},
\end{equation}
where
\[
|\psi^{\perp(k)}_n\rangle
=
P_n^{(k)}(\tilde H)|\Phi^{(k)} \rangle, \qquad n=0,\ldots,n^{\rm max}.
\]
This is classically solved to obtain an  approximate ground state $|\Psi^{(k)}_0\rangle$ then converted into a quantum state.
The latter is used to initialize a new OQKD loop as, 
\begin{align}
& |\Phi^{(k+1)} \rangle \leftarrow  |\Psi^{(k)}_0\rangle \\
& k \leftarrow k+1. 
\end{align}
The algorithm stops when convergence criteria are reached.

More specifically, at restart \(k\), since each Lanczos polynomial admits a Chebyshev expansion
\begin{equation}
    P_n^{(k)}(x) = \sum_{m=0}^{n} c_{nm}^{(k)}T_m(x),
\end{equation}
substituting this expansion into the development of the estimated ground-state in the OQKD basis
\begin{equation}
|\Psi_0^{(k)} \rangle = \sum_{n=0}^{n_{\rm max}} a_n ^{(k)} |\psi^{\perp(k)}_n\rangle
\end{equation}
leads to
\begin{equation}
    |\Psi^{(k+1)}_0\rangle = \sum_m g_m^{(k)} T_m(\tilde H)|\Psi^{(k)}_0\rangle,
\end{equation}
where
\[
g_m^{(k)}
=
\sum_{n=0}^{n_{\rm max}}
a_n^{(k)}
c_{nm}^{(k)}.
\]
The approximate ground state can be written iteratively as
\[
|\Psi^{(k)}_0\rangle
=
Q^{(k)}(\tilde H)
|\Psi^{(k-1)}_0\rangle,
\]
with
\[
Q^{(k)}(x)
=
\sum_m
g_m^{(k)}
T_m(x).
\]
Note that the coefficients $a_n^{(k)}$ modulate the amplitudes of $g_m^{(k)}$ since by construction $\sum_{n=0}^{n_{\rm max}} |a_n^{(k)}|^2 = 1$, decreasing the norm of $Q^{(k)}(x)$. 
After $R$ restart cycles, the approximate ground state can be written as
\begin{equation}
|\Psi^{(R)}_0\rangle=Q^{(R)}(\tilde H)\,
|\Phi_0\rangle,
\label{eq:intaitlstate_poly}
\end{equation}where
\begin{equation}
Q^{(R)}(x)
=
Q^{(R-1)}(x)
Q^{(R-2)}(x)
\cdots
Q^{(1)}(x)
Q^{(0)}(x),
\end{equation}The success probability associated with implementing the accumulated polynomial $Q^{(R)}$ in a single GQSP circuit is determined by the normalization condition required for its realization within the GQSP framework.
\begin{equation}
\Lambda^{(R)}
=
\max_{|z|=1}
\left|
Q^{(R)}(z)
\right|,
\label{eq:lambda_restart}
\end{equation}
The ROQKD algorithm is summarized in Algorithm~\ref{alg:restart_oqkd}.

\begin{algorithm}[ht]
\caption{Restarted OQKD (ROQKD)}
\label{alg:restart_oqkd}
\begin{enumerate}
\item \textbf{Input:}
\[
\tilde H,\quad
|\Phi^{(0)}_0\rangle,\quad
n^\text{max},\quad
R,
\]
where \(n^{\rm max}\) is the fixed OQKD depth and \(R\) is the number of restart cycles.

\item \textbf{Restart iterations:}
For \(k=0,\ldots,R-1\):
\begin{enumerate}
\item \label{step:OKQD} Use OQKD described in Algorithms~\ref{app:quantum_algo} with initial state
\(
|\Phi^{(k)}_0\rangle.
\)
Obtain  the approximate ground-state coefficients
\[
a^{(k)}
=
(a^{(k)}_0,\ldots,a^{(k)}_{n^{\rm max}}).
\]
\item Expand the approximate ground state in the Chebyshev basis,
\[
|\Psi^{(k)}_0\rangle
=
Q^{(k)}(\tilde H)
|\Phi^{(k-1)}_0\rangle,
\]
where
\[
Q^{(k)}(x)
=
\sum_m
g^{(k)}_m
T_m(x).
\]
\item Construct a GQSP circuit implementing
\(
Q^{(k)}(\tilde H)
\)
and prepare the new initial state according to Eq.~(\ref{eq:intaitlstate_poly}).
\[
|\Phi^{(k+1)}_0\rangle \leftarrow |\Psi^{(k)}_0\rangle.
\]
\end{enumerate}
\item \textbf{Output:}
Approximate ground state
\[
|\psi^{(R)}_0\rangle.
\]
\end{enumerate}
\end{algorithm}
Algorithm~\ref{alg:restart_oqkd} can be naturally extended to other variants of Quantum Krylov methods by replacing step~(\ref{step:OKQD}) with the corresponding subspace-construction routine. For instance, in the case of CQKD, step~(\ref{step:OKQD}) is replaced by the CQKD procedure, which measures the projected Hamiltonian and the corresponding overlap matrix in the non-orthonormal Chebyshev basis and solves the resulting generalized eigenvalue problem to obtain the ground-state coefficients
\[
(a^{(k)}_0,\ldots,a^{(k)}_{n^\text{max}}).
\]

In contrast to OQKD, the Chebyshev basis vectors must be prepared at every restart. While this preparation does not introduce an exponentially decreasing probability of success, it imposes a practical limitation on the maximum Krylov subspace dimension that can be reached at each restart. Throughout this work, we therefore choose the subspace dimension $n_k$ at each restart as the largest value for which the generalized eigenvalue problem can still be solved reliably without introducing thresholding. We note that choosing the same fixed subspace dimension at every restart leads to the same convergence behavior as restarted OQKD, provided that this dimension is numerically accessible.

More generally, Algorithm~\ref{alg:restart_oqkd} provides a generic restarted Quantum Krylov framework in which only the subspace-construction routine is modified. The choice of this routine can be optimized according to the target quantum hardware, available resources, and numerical stability considerations, while the restart strategy itself remains unchanged.

\subsection{Numerical efficiency and state preparation}
\begin{figure*}[t]
    \centering
    \includegraphics[width=\textwidth]{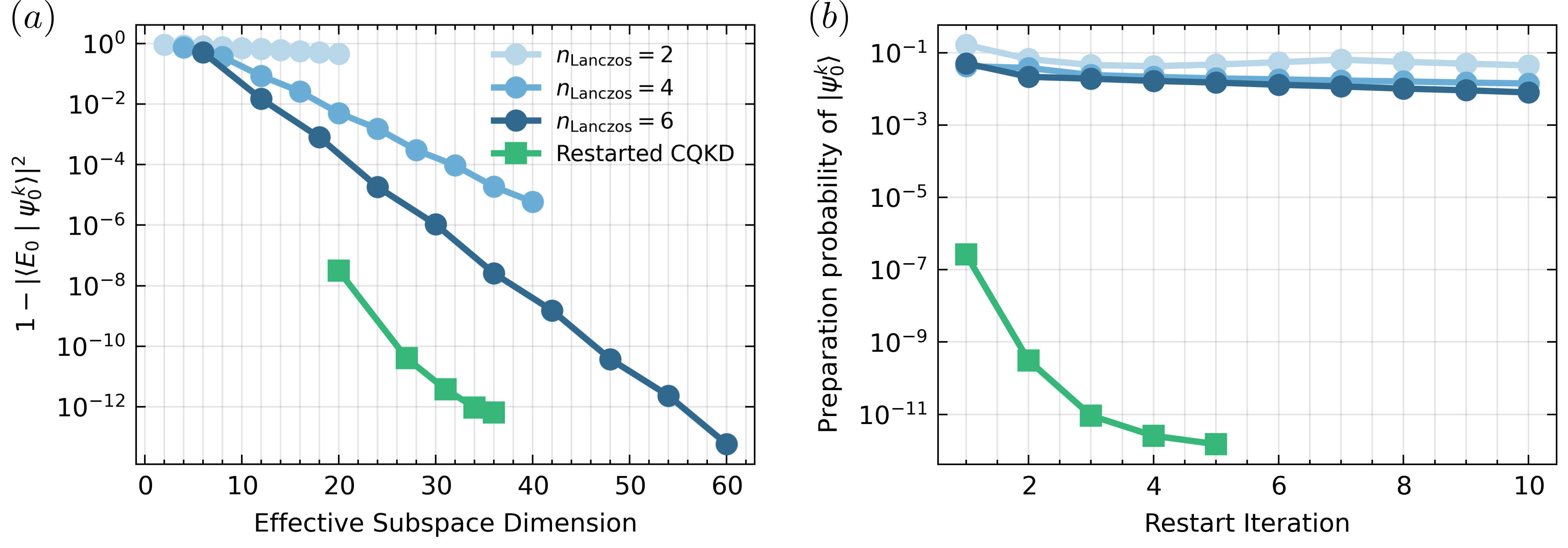}
    \caption{Performance of the restarted OQKD state-preparation protocol. (a) Infidelity,
    $1-\left|\langle E_0|\psi_0^{(k)}\rangle\right|^2$, of the prepared state with respect to the exact ground state as a function of the restart iteration. Despite the nearly constant state-preparation probability, the infidelity decreases monotonically toward zero, demonstrating that the restarted protocol progressively improves the ground-state approximation while maintaining a favorable implementation cost. (b) Probability of successfully preparing the approximate ground state after each restart iteration from the original easily prepared initial state by implementing the accumulated GQSP polynomial of Eq.~(\ref{eq:intaitlstate_poly}). The success probability remains nearly constant throughout the restart procedure. Both panels are shown on a logarithmic scale.}
    \label{fig:Fig_ROQKD}
\end{figure*}

The numerical performance of the restarted protocols for different Quantum Krylov subspace diagonalization schemes is presented in Fig.~\ref{fig:Fig_ROQKD}. Figure~\ref{fig:Fig_ROQKD}(a) compares the infidelity obtained with ROQKD for different Lanczos subspace dimensions, $n_{\rm Lanczos}$, and with RCQKD, where the Krylov subspace dimension at each restart is chosen as the largest value that allows a stable solution of the generalized eigenvalue problem without thresholding, as discussed above.

Importantly the restarted OQKD protocol inherits the exponential convergence of the Lanczos method. Moreover, the convergence rate increases with the maximum Lanczos subspace dimension, $n_{\rm Lanczos}$, leading to a steeper exponential decay of the infidelity. In other words, larger Krylov subspaces require fewer restart cycles to reach a given target accuracy. The restarted CQKD protocol exhibits a comparable convergence behavior while operating at the largest numerically stable Krylov dimension accessible without thresholding.
Figure~\ref{fig:Fig_ROQKD}(b) reports the success probability associated with preparing the approximate ground state at one restart iteration, which subsequently serves as the initial state for the next restart. For the ROQKD protocol, the success probability remains nearly constant throughout the restart procedure, even as the number of restart iterations increases. In contrast, restarted CQKD (RCQKD) exhibits an exponential decrease in the success probability. This behavior originates from the fact that each restart already employs the largest numerically stable Krylov subspace, making the corresponding polynomial significantly more difficult to implement through GQSP.

It is important to emphasize that the exponentially small success probabilities observed for RCQKD, as well as for the preparation of the Lanczos vectors in OQKD (see Fig.~\ref{fig:P_succ}), are a direct consequence of the orthogonalization and normalization procedures. These procedures lead to rapidly increasing coefficients in the Chebyshev polynomial representation, which in turn require increasingly stronger normalization to satisfy the GQSP implementation conditions, thereby yielding an exponentially decreasing probability of success.

A systematic study of the interplay between the polynomial growth associated with the orthogonalization procedure and the convergence of the restarted CQKD and OQKD protocols as the system size increases is beyond the scope of the present work and constitutes an interesting direction for future research.

 \section{Conclusion}

In this work, we introduced an Orthogonal Quantum Krylov Diagonalization (OQKD) framework that faithfully reproduces the structure and convergence properties of the classical Lanczos algorithm on a quantum computer. By reformulating the Lanczos recursion at the operator level, OQKD generates orthogonal Krylov vectors through polynomial transformations of the Hamiltonian, thereby preserving the orthogonality and tridiagonal structure of the classical Lanczos basis while completely eliminating the overlap-matrix regularization required in existing quantum Krylov methods. Numerical simulations of the $J_1$--$J_2$ Heisenberg model demonstrate convergence behavior essentially identical to that of the classical Lanczos algorithm, reaching machine precision while maintaining a well-conditioned projected basis. Furthermore, the Lanczos polynomials can be implemented using block encoding and GQSP with the same asymptotic query complexity as Chebyshev-based quantum Krylov methods.

We further showed that OQKD removes the overlap-conditioning bottleneck that governs the measurement complexity of conventional QKD and CQKD. Since the method completely avoids overlap-matrix regularization, its measurement complexity is independent of the overlap between the initial state and the target eigenstate. The remaining practical limitation is instead the implementation of high-degree polynomial transformations within the GQSP framework, whose success probability decreases with the Lanczos dimension.

Beyond its role as an orthogonal quantum eigensolver, OQKD naturally enables a practical state-preparation protocol through a restarted implementation. Rather than applying a single high-degree Lanczos polynomial, the approximate ground state obtained after each OQKD run is used as the initial state for the next, resulting in a sequence of fixed low-degree polynomial transformations. At every restart, the approximate ground state can be expressed as an explicit GQSP polynomial acting on the original easily prepared initial state, as given by Eq.~(\ref{eq:intaitlstate_poly}). Consequently, the quantum circuit required to prepare the approximate ground state after each restart from the original easily prepared initial state is known explicitly, while the probability of successfully implementing this state-preparation circuit remains nearly constant throughout the restart procedure.

Overall, this work establishes OQKD as both an orthogonal quantum analog of the classical Lanczos algorithm and a practical framework for quantum state preparation. By bridging the gap between the classical orthogonal Lanczos method and quantum subspace-diagonalization techniques, while naturally integrating with block encoding and GQSP, the proposed framework provides a scalable route toward robust quantum eigensolvers and efficient state-preparation protocols for Quantum Phase Estimation. More broadly, it opens a promising path toward scalable quantum algorithms for large-scale, industrially relevant eigenvalue problems.

\section*{Acknowledgments}
This work was supported in part by the Maison du Quantique de Nouvelle-Aquitaine “HybQuant”, as 	part of the HQI initiative and France 2030, under the French National Research Agency (ANR) 	grant ANR-22-PNCQ-0002 and the PEPR Quantique under Grant No. ANR-23-PETQ-0006. We also thanks the EUR Light S$\&$T (PIA3 Program, Grant No. ANR-17-EURE-0027)

\section*{Authors contribution}
Algorithm implementation and data acquisition: H.R. (lead) A.P. and O.L. (support); Concept and methodology: J.M., M.S. (lead) and H.R., A.P. and O.L.  (support); Project management: J.M. and M.S; Supervision: M.S. (lead) and C.D. (support); Writing original draft: H.R.. All authors edited and commented on the manuscript.

\bibliographystyle{plain}
\bibliographystyle{apsrev4-2}  
\bibliography{Biblio}


\appendix 

\section{Recursive determination of the Lanczos coefficients}

We now derive the recursive relation used to determine the normalization coefficients $\beta_{k+1}$ without explicitly constructing or storing the full Lanczos basis. Starting from the Lanczos recurrence relation,
\begin{equation}
H|\psi_k\rangle
=
\beta_{k+1}|\psi_{k+1}\rangle
+
\alpha_k|\psi_k\rangle
+
\beta_k|\psi_{k-1}\rangle ,
\label{eq:lanczos_recurrence_beta_derivation}
\end{equation}
the action of the Hamiltonian on $|\psi_k\rangle$ decomposes into three mutually orthogonal directions:
$|\psi_{k-1}\rangle$, $|\psi_k\rangle$, and $|\psi_{k+1}\rangle$. The coefficients of the first two components are already determined by previous iterations, while the norm of the remaining orthogonal component fixes $\beta_{k+1}$.

Rearranging Eq.~\eqref{eq:lanczos_recurrence_beta_derivation}, one obtains
\begin{equation}
\beta_{k+1}|\psi_{k+1}\rangle
=
(H-\alpha_k)|\psi_k\rangle
-
\beta_k|\psi_{k-1}\rangle .
\label{eq:isolate_next_lanczos_vector}
\end{equation}
Taking the squared norm of both sides gives
\begin{align}
\beta_{k+1}^2
&=
\left\|
(H-\alpha_k)|\psi_k\rangle
-
\beta_k|\psi_{k-1}\rangle
\right\|^2
\nonumber\\
&=
\langle\psi_k|(H-\alpha_k)^2|\psi_k\rangle
+\beta_k^2
\nonumber\\
&\quad
-
2\beta_k
\operatorname{Re}
\left[
\langle\psi_{k-1}|
(H-\alpha_k)
|\psi_k\rangle
\right].
\label{eq:beta_norm_expansion}
\end{align}
Using the Lanczos orthogonality relations,
\begin{equation}
\langle\psi_{k-1}|\psi_k\rangle=0,
\qquad
\langle\psi_{k-1}|H|\psi_k\rangle=\beta_k,
\label{eq:lanczos_orthogonality_relations}
\end{equation}
the cross term in Eq.~\eqref{eq:beta_norm_expansion} reduces to $2\beta_k^2$. Therefore,
\begin{equation}
\beta_{k+1}^2
=
\langle\psi_k|(H-\alpha_k)^2|\psi_k\rangle
-
\beta_k^2 .
\label{eq:beta_before_expansion}
\end{equation}
Finally, using
\begin{equation}
\alpha_k=\langle\psi_k|H|\psi_k\rangle,
\label{eq:alpha_definition}
\end{equation}
and expanding $(H-\alpha_k)^2$, we obtain
\begin{align}
\beta_{k+1}^2
&=
\langle\psi_k|H^2|\psi_k\rangle
-2\alpha_k\langle\psi_k|H|\psi_k\rangle
+\alpha_k^2
-\beta_k^2
\nonumber\\
&=
\langle\psi_k|H^2|\psi_k\rangle
-\alpha_k^2
-\beta_k^2 .
\label{eq:beta_recursive_relation}
\end{align}
Thus, the normalization coefficient is determined recursively as
\begin{equation}
\beta_{k+1}
=
\sqrt{
\langle\psi_k|H^2|\psi_k\rangle
-\alpha_k^2
-\beta_k^2
}.
\label{eq:beta_recursive_final}
\end{equation}

For the rescaled Hamiltonian $\tilde H$, the same relation reads
\begin{equation}
\tilde\beta_{k+1}^2
=
\langle\psi_k|\tilde H^2|\psi_k\rangle
-\tilde\alpha_k^2
-\tilde\beta_k^2 .
\label{eq:beta_recursive_rescaled}
\end{equation}
This relation provides a fully recursive determination of the Lanczos normalization coefficients without requiring explicit orthogonalization against the full Krylov basis.

Once the coefficients $\tilde\alpha_k$ and $\tilde\beta_k$ are known, the Lanczos polynomials can be generated recursively. The Lanczos vectors may then be written as polynomial transformations of the rescaled Hamiltonian acting on the initial state,
\begin{equation}
|\psi_n\rangle
=
P_n(\tilde H)|\Phi_0\rangle,
\label{eq:lanczos_vector_polynomial_form}
\end{equation}
where $P_n(x)$ denotes the $n$-th Lanczos polynomial defined by the recurrence relation associated with the coefficients $\tilde\alpha_k$ and $\tilde\beta_k$. This polynomial formulation naturally enables quantum implementations based on block encoding and Chebyshev polynomial expansions, as discussed in the following appendix.

\label{app:derivation}

\section{Chebyshev representation of the Lanczos polynomials}
\label{app:C_coefficient_algo}

The Lanczos polynomials satisfy the three-term recurrence relation

\begin{equation}
P_{n+1}(x)
=
\frac{
(x-\alpha_n)P_n(x)
-
\beta_n P_{n-1}(x)
}{
\beta_{n+1}
},
\label{eq:lanczos_recursion_appendix}
\end{equation}

with

\begin{equation}
P_0(x)=1,
\qquad
P_1(x)=\frac{x-\alpha_0}{\beta_1}.
\end{equation}

To obtain a quantum implementation, it is convenient to expand the Lanczos polynomials in the Chebyshev basis,

\begin{equation}
P_n(x)
=
\sum_{m=0}^{n}
c_{nm}T_m(x).
\label{eq:lanczos_cheb_appendix}
\end{equation}

The Chebyshev basis is particularly useful because multiplication by \(x\) obeys

\begin{equation}
xT_m(x)
=
\frac12
\left[
T_{m+1}(x)
+
T_{m-1}(x)
\right].
\label{eq:cheb_recurrence_appendix}
\end{equation}
Substituting the Chebyshev expansion

\begin{equation}
P_n(x)
=
\sum_{m=0}^{n}
c_{nm}T_m(x)
\end{equation}
into the Lanczos recurrence relation
\begin{equation}
P_{n+1}(x)
=
\frac{
(x-\alpha_n)P_n(x)
-
\beta_n P_{n-1}(x)
}{
\beta_{n+1}
},
\end{equation}
and using Eq.~\eqref{eq:cheb_recurrence_appendix} gives

\begin{align}
P_{n+1}(x)
=
\frac{1}{\beta_{n+1}}
\Bigg[
&
\frac12
\sum_m
c_{nm}
\left(
T_{m+1}(x)
+
T_{m-1}(x)
\right)
\nonumber\\
&
-\alpha_n
\sum_m
c_{nm}T_m(x)
-
\beta_n
\sum_m
c_{n-1,m}T_m(x)
\Bigg].
\end{align}

Collecting equal Chebyshev orders yields the coefficient recursion

\begin{equation}
c_{n+1,m}
=
\frac{
\frac12 c_{n,m-1}
+
\frac12 c_{n,m+1}
-
\alpha_n c_{n,m}
-
\beta_n c_{n-1,m}
}{
\beta_{n+1}
},
\label{eq:cheb_coeff_recursion}
\end{equation}with the convention that coefficients outside the polynomial degree vanish.

Starting from the coefficients of

\begin{equation}
P_0(x)=1,
\qquad
P_1(x)=\frac{x-\alpha_0}{\beta_1},
\end{equation}namely

\begin{equation}
c_{0,0}=1,
\qquad
c_{1,0}=-\frac{\alpha_0}{\beta_1},
\qquad
c_{1,1}=\frac{1}{\beta_1},
\end{equation}all higher-order coefficients \(c_{nm}\) can be generated recursively from Eq.~\eqref{eq:cheb_coeff_recursion}. Consequently, the complete Chebyshev representation of the Lanczos polynomials can be constructed using only the Lanczos coefficients \(\{\alpha_n,\beta_n\}\).


\section{From GQSP to a polynomial of the Hamiltonian}
\label{App:GQSP}

We consider the notations of section \ref{parag:impelntation}.
Let
\begin{equation}
\tilde H|\lambda_i\rangle
=
\lambda_i |\lambda_i\rangle .
\end{equation}

\begin{equation}
W_i
=
\begin{pmatrix}
\cos\theta_i & \sin\theta_i\\
-\sin\theta_i & \cos\theta_i
\end{pmatrix},
\qquad
\cos\theta_i=\lambda_i.
\end{equation}

The full walk operator therefore decomposes as

\begin{equation}
W
=
\bigoplus_i W_i.
\end{equation}
To implement a polynomial transformation of the Hermitian operator $\tilde H$ on a quantum computer, the GQSP framework can be used, giving:
\begin{equation}
(\langle 0|_a\otimes I_s)\,
\text{GQSP}_n[W]\,
(|0\rangle_a\otimes I_s)
=
\frac{1}{\Lambda_n}P_n(\tilde H),
\label{eq:block_encoding2}
\end{equation}
where $\Lambda_n$ is a normalization factor 
defined by
\begin{equation}
\Lambda_n\ge \max_{|z|=1}|p_n(z)|,
\quad p_n(z)=\sum_{k=0}^n c_{nk} z^k.
\label{eq:block_encoding3}
\end{equation}
The GQSP sequence consists of alternating applications of the walk operator and rotations,
\begin{equation}
\mathrm{GQSP}_n[W]
=
R_nWR_{n-1}W\cdots WR_1WR_0,
\label{eq:gqsp_sequence}
\end{equation}
where each rotation \(R_j\) depends on parameters
\(\{\theta_j,\phi_j,\lambda\}\) and acts on the signal-processing ancilla qubit. 
As demonstrated in Ref.~\cite{motlagh2024gqsp}, the GQSP sequence is synthesized so that the normalized Laurent polynomial
\begin{equation}
\tilde p_n(z)
=
\frac{1}{\Lambda_n}
\sum_{k=0}^{n}
c_{nk} z^k
,
\label{eq:lanczos_beard_poly}
\end{equation}
is applied independently within each invariant subspace,
\begin{equation}
\mathrm{GQSP}_n[W]
|\Lambda_i^\pm\rangle
=
\tilde p_n(z)
(e^{\pm i\theta_i})
|\Lambda_i^\pm\rangle .
\label{eq:ciruit}
\end{equation}

Projecting the transformed state back onto the encoded subspace gives
\begin{align}
&
\left(
\langle0|_a
\langle\lambda_i|
\right)
\mathrm{GQSP}_n[W]
\left(
|0\rangle_a
|\lambda_i\rangle
\right)
\nonumber\\
&=
\frac12
\left[
\tilde p_n(z)
(e^{i\theta_i})
+
\tilde p_n(z)
(e^{-i\theta_i})
\right]
\nonumber\\
&=
\frac12
\sum_{k=0}^{n}
\frac{c_{nk}}{\Lambda_n}
\left(
e^{ik\theta_i}
+
e^{-ik\theta_i}
\right).
\label{eq:eigenphase_appendix}
\end{align}
Using the Chebyshev identity

\begin{equation}
T_k(\cos\theta_i)
=
\cos(k\theta_i)
=
\frac{
e^{ik\theta_i}
+
e^{-ik\theta_i}
}{2},
\end{equation}
together with Eq.~(\ref{eq:eigenphase_appendix}), yields

\begin{equation}
\frac{1}{\Lambda_n}
\sum_{k=0}^{n}
c_{nk}
T_k(\lambda_i)
=
\frac{1}{\Lambda_n}
P_n(\lambda_i).
\end{equation}
Since this relation holds for every eigenvalue \(\lambda_i\), the spectral decomposition of \(\tilde H\) implies

\begin{equation}
(\langle0|_a\otimes I)
\,\mathrm{GQSP}_n[W]\,
(|0\rangle_a\otimes I)
=
\frac{1}{\Lambda_n}
P_n(\tilde H).
\end{equation}

Thus, the qubitized GQSP construction realizes the desired Lanczos polynomial transformation of the Hamiltonian.
\subsection{Calculation of the GQSP rotation angles}
\label{app:angels}

Given the normalized polynomial \(\widetilde p_n(z)\), the corresponding GQSP circuit is synthesized by first constructing a complementary polynomial \(Q(z)\), 
\begin{equation}
Q(z)
=
\sum_{k=0}^{n}
q_k z^k,
\end{equation}
satisfying

\begin{equation}
|\widetilde p_n(z)|^2
+
|Q(z)|^2
=
1,
\qquad
|z|=1.
\label{eq:complementary_condition_appendix}
\end{equation}
The existence of the complementary polynomial \(Q(z)\) follows from the spectral factorization condition

\begin{equation}
Q(z)Q^{\ast}(1/z^{\ast})
=
1
-
\widetilde p_n(z)
\widetilde p_n^{\ast}(1/z^{\ast}).
\label{eq:spectral_factorization}
\end{equation}Thus, \(Q(z)\) may in principle be obtained by factorizing the roots of

\begin{equation}
F(z)
=
1
-
\widetilde p_n(z)
\widetilde p_n^{\ast}(1/z^{\ast}).
\end{equation}However, rather than performing an explicit spectral factorization, Ref.~\cite{motlagh2024gqsp} introduces an efficient numerical optimization procedure that directly computes the coefficients of a complementary polynomial \(Q(z)\) satisfying

\begin{equation}
|\widetilde p_n(z)|^2
+
|Q(z)|^2
=
1,
\qquad
|z|=1.
\end{equation}

Once the polynomial pair \((\widetilde p_n,Q)\) has been obtained, the GQSP rotation angles
\(
\{\theta_j,\phi_j,\lambda\}
\)
are extracted using the recursive phase-factor decomposition of Ref.~\cite{motlagh2024gqsp}. Let

\begin{equation}
\widetilde p_n(z)
=
\sum_{k=0}^{n}
a_k z^k,
\qquad
Q(z)
=
\sum_{k=0}^{n}
b_k z^k,
\end{equation}and denote by \(a_n\) and \(b_n\) the highest-degree coefficients. The final rotation angles are determined from

\begin{equation}
\theta_n
=
\tan^{-1}
\left(
\frac{|b_n|}{|a_n|}
\right),
\end{equation}

and

\begin{equation}
\phi_n
=
\mathrm{Arg}
\left(
\frac{a_n}{b_n}
\right).
\end{equation}

These values are chosen such that the highest-degree terms cancel when the inverse of the final GQSP layer is applied, producing a reduced polynomial pair of degree \(n-1\). The same procedure is then repeated recursively for degrees \(n-1,n-2,\ldots,0\) until constant polynomials are obtained.

The recursive decomposition therefore determines the angle pairs
\(
(\theta_j,\phi_j)
\)
by successively removing the highest-degree polynomial contribution. Once the recursion reaches degree zero, the remaining constant complementary polynomial
\(
Q^{(0)}(z)=b_0
\)
determines the global phase parameter
\begin{equation}
\lambda=\mathrm{Arg}(b_0).
\end{equation}

Together,
\(
\{(\theta_j,\phi_j)\}_{j=0}^{n}
\)
and \(\lambda\) completely specify the GQSP circuit. The resulting angles define the GQSP sequence

\begin{equation}
R_nWR_{n-1}W\cdots WR_1WR_0,
\end{equation}
which implements the desired polynomial transformation of the qubitized walk operator~\cite{motlagh2024gqsp}.


\section{OQKD as state preparation}

In the context of QPE, the OQKD approach can be employed as an efficient initial-state preparation scheme. By producing trial states with significantly enhanced overlap with the exact ground state, OQKD alleviates the catastrophic orthogonality problem and thereby improves the success probability of QPE.

To estimate the number of iterations required, and consequently the depth of the quantum circuit, we rely on the Kaniel--Paige (KP) theorem, which provides an upper bound on the error as a function of the spectral properties of the system and the initial-state overlap:
\begin{equation}
\mathcal{E}(D)
\leq
f(\gamma_0^2)
\left(\frac{1}{R^2}\right)^D,
\label{eq:KPST}
\end{equation}
where
\[
f(\gamma_0^2)=\frac{1-\gamma_0^2}{\gamma_0^2},
\]
and \(\gamma_0\) denotes the overlap amplitude between the initial state \(|\Phi_0\rangle\) and the exact ground state \(|E_0\rangle\),
\[
\gamma_0 = \langle E_0 | \Phi_0\rangle .
\]
The quantity \(R\) is defined as
\begin{equation}
    R=\frac{1+r+2\sqrt{r}}{1-r},
\end{equation}
with
\begin{equation}
r=
\frac{\lambda_2-\lambda_1}{\lambda_n-\lambda_1},
\qquad
0\le r\le1,
\label{eq:r}
\end{equation}
where \(\lambda_1\) is the ground-state energy, \(\lambda_2\) the first excited-state energy, and \(\lambda_n\) the largest eigenvalue of the Hamiltonian.

In the thermodynamic limit of large system size \(N\), the overlap factor can be approximated as
\begin{equation}
f(\gamma_0^2)
\simeq
\frac{1}{\gamma_0^2}.
\end{equation}

For spin systems, the ground-state energy typically scales extensively with the system size \(N\). Assuming that the total bandwidth \(\lambda_n-\lambda_1\) also scales linearly with \(N\), while the first excitation gap behaves as
\begin{equation}
    \lambda_2-\lambda_1
\sim
\frac{1}{N},
\end{equation}

one obtains
\begin{equation}
r
\sim
\frac{1/N}{N}
=
\frac{1}{N^2}.
\end{equation}

Expanding \(R^2\) in the large-\(N\) limit then gives
\begin{equation}
R^2
\simeq
1+\frac{4}{N}
+
\mathcal{O}\!\left(\frac{1}{N^2}\right).
\end{equation}

Moreover, if one assumes a completely random initial state in the computational basis, the typical overlap with the exact ground state scales as
\begin{equation}
\gamma_0^2
\sim
\frac{1}{2^N},
\end{equation}
since the Hilbert-space dimension is \(2^N\).

Substituting these asymptotic scalings into Eq.~\ref{eq:KPST} yields
\begin{equation}
\mathcal{E}(D)
\leq
2^N
\left(\frac{1}{1+\frac{4}{N}}\right)^D .
\end{equation}

Using
\[
\log(1+x)\simeq x,
\qquad
(x\ll1),
\]
we obtain
\begin{equation}
\log \mathcal{E}(D)
\leq
N\log 2
-
D\log\left(1+\frac{4}{N}\right)
\simeq
N\log 2
-
\frac{4D}{N}.
\end{equation}

Requiring a fixed target accuracy \(\varepsilon(D)\) therefore implies
\begin{equation}
D
\sim
\frac{N^2}{4}\log 2,
\end{equation}
such that the number of Lanczos iterations scales asymptotically as
\begin{equation}
D=\mathcal{O}(N^2).
\end{equation}
\label{app:saling_Eq}

Under the physically motivated assumptions that the spectral bandwidth scales as \(O(N)\), the spectral gap as \(O(1/N)\), and the squared overlap of the initial state with the target ground state as \(O(2^{-N})\), which are generally expected in the thermodynamic limit, the KP bound predicts a required Lanczos subspace dimension scaling as \(D = O(N^2)\).

\medskip
\noindent\textbf{Success Probability of the State-Preparation Protocol.}
\label{app:state_preparation_probability}
\begin{figure}
    \centering
    \includegraphics[width=1\linewidth]{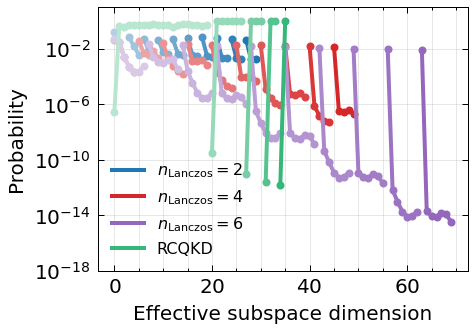}
    \caption{\label{fig:state_preparation_probability}
Success probability for preparing the Krylov basis vectors in the restarted OQKD (ROQKD) protocol as a function of the effective subspace dimension for different maximum Lanczos subspace dimensions, $n_{\rm Lanczos}=2$, $4$, and $6$. For the restarted CQKD (RCQKD) protocol, the Krylov subspace dimension at each restart is chosen according to the protocol described above, namely as the largest numerically stable value that avoids thresholding. Darker shades correspond to successive restart iterations.}
    \label{fig:placeholder}
\end{figure}

The success probabilities for preparing each Krylov subspace vector are shown in Fig.~\ref{fig:state_preparation_probability}. In ROQKD, the probability of preparing the restarted initial state remains nearly constant throughout the restart procedure and is significantly larger than the probability of preparing the Lanczos vectors generated within each Krylov subspace. Consequently, the exponential decrease in the preparation probability of the Lanczos vectors does not prevent successive restart iterations from being implemented efficiently.

In contrast, the behavior of RCQKD is reversed. As discussed above, the probability of preparing the restarted initial state is already very small and decreases exponentially with the restart iteration. This behavior originates from the accumulated orthogonalization polynomial required to reconstruct the approximate ground state. However, once the restarted initial state has been successfully prepared, the probability of generating the corresponding Chebyshev basis vectors remains close to unity throughout the Krylov subspace. Therefore, in RCQKD the exponential cost is associated with preparing the restarted state itself, whereas in ROQKD it is associated with preparing the Lanczos basis vectors within each restart.

\end{document}